# How Bohr's Copenhagen interpretation is realist and solves the measurement problem.

Govind Krishnan.V

July 2023

## Abstract

The field of interpretations of quantum mechanics emerged in an attempt to solve the measurement problem. This turned on the perception that Niels Bohr avoided addressing the measurement problem by taking an instrumentalist view of quantum mechanics. I argue that this perception is mistaken and Bohr's interpretation of quantum mechanics is realist. Moreover, Bohr's interpretation, which is different from textbook quantum mechanics, (which is due more to Von Neumann and Paul Dirac), succeeds in solving the measurement problem. While the claim that Bohr 'dissolves' the measurement problem within the limits of the epistemological framework he assumes, has been made by a few authors, rarely has the case been made that Bohr's project unambiguously and completely overcomes the measurement problem. I make the strong case that Bohr eliminated the measurement problem altogether. For this, I put forward two new postulates through which to make sense of Bohr's interpretation. The article thus seeks to single out Bohr's interpretation from the various views that together go under the umbrella of orthodox quantum mechanics, and which have been traditionally considered susceptible to the measurement problem. It shows that Bohr's interpretation should be classified along with those like hidden variable theories, collapse models, modal interpretations etc., which offer a solution to the measurement problem and are committed to a realist ontology.

## 1. Introduction

Even 95 years after the fifth Solvay conference, confusion reigns about what exactly Niels Bohr's Copenhagen interpretation says. He continues to be regarded as a positivist (See Oldofredi and Esfeld) (2019), an idealist[1], an instrumentalist[2], and even as a proponent of the shut up-and-calculate approach. This confusion attends not only Bohr's epistemology, but even the content of the theory. It is not uncommon for physicists to mistakenly attribute the Dirac-Von Neumann formalism of wave collapse to Bohr's Copenhagen interpretation. While this conflation is less prevalent among physicists and philosophers of physics concerned with the foundations of quantum mechanics; Bohr's interpretation still has a bad name with many working in quantum foundations. Many see his effort as an attempt to avoid foundational questions in quantum mechanics, including questions about the reality of the wave function, the measurement problem and the physical meaning of quantum theory. He is also seen as an obscurantist, who deliberately tried to obfuscate philosophical and foundational questions about quantum theory. Durr and Lazarovici (2019) write *"We can also put it succinctly like*

---

[1] The charge of idealism against the Copenhagen interpretation was more prominent during the early decades of its formulation. Though less encountered now, such an understanding has by no means disappeared. For a recent opinion of Bohr's interpretation as idealistic, see Barret (2019)

[2] Bohr is most commonly understood as taking an instrumentalist view of quantum mechanics. See Harrigan and Spekkens (2010) and Bacciagaluppi (2013).

*this: the Copenhagen interpretation of quantum mechanics is not physics, but a mixture of a mathematical formalism that successfully describes measurement statistics, and psychological warfare against a better understanding."*

The contribution of this article is the following: it shows that Niels Bohr's version of the Copenhagen interpretation solves the measurement problem without abandoning realism or resorting to instrumentalism. It also shows that Bohr's theory can be accommodated within scientific realism, if the latter doctrine is construed more broadly than it usually is. These arguments are novel and do not appear anywhere in the literature of quantum foundations, as far as the author is aware. If these claims hold true, as I endeavour to show, it has wide ranging implications for the quantum foundations community as well as the philosophy of physics community at large. One major implication is that Bohr's Copenhagen interpretation would have to be considered as a candidate for the true description of quantum reality along with theories such as Bohmian mechanics, many worlds interpretation, collapse models, modal interpretations etc. This would be a significant shift from the current consensus.
I present two new postulates about the Copenhagen interpretation which are required in order to give the theory complete coherence and explanatory capacity. The shadows of both are present in Bohr's work, though he doesn't state them explicitly, nor do any of the philosophers who have written on Bohr. But these postulates are not only hinted at in Bohr's work, they are both logical necessities if Bohr's interpretation is to be conceptually free of all ambiguities. Taken together with Bohr's epistemological framework, the two postulates A and B solve the measurement problem.

I will briefly introduce the postulates here, before giving a detailed explanation later.

A) The Copenhagen interpretation consists of two different modes of description, a symbolic description, and a representative description. The description of a quantum system through the wave function and through the experimental results of a measurement, are complementary and mutually exclusive descriptions. The first is a symbolic description, while the second is a representative description.

B) The measurement apparatus is the epistemic condition of possibility of access to quantum systems.

Postulate A provides a generalised theoretical lens through which to view Bohr's interpretation and give it maximum coherence. It also helps to understand in a perspicuous way, why one aspect of the measurement problem, the problem of collapse, does not arise for the Copenhagen interpretation.
Postulate B is an epistemological generalisation concerning the role of the measurement apparatus in Bohr's theory. It is the foundational epistemological principle that makes both Postulate A and Bohr's pragmatised Kantian epistemology (On Bohr's pragmatised Kantianism, See Folse 1994) fully coherent in their application to quantum mechanics. It also dissolves the observer dependence of measurement, which orthodox quantum mechanics has never been able to fully exorcise.

While the solution to the measurement problem is the main focus of this article, I will deal with it only in the later sections. In order to make sense of how Bohr solves the measurement problem, it is necessary to obtain a clear conceptual understanding of Bohr's Copenhagen interpretation. This is not an easy task. What Bohr proposes is a fundamental reorganisation of the epistemological foundation of physics. Also, Bohr's interpretation has been misunderstood and consequently criticised from such a variety of positions, that one has to clear up the cloud of confusion surrounding the Copenhagen interpretation before proceeding further. Thus, a major part of the article leading up to the solution of the measurement problem would be expository in nature, elucidating the nature of Bohr's interpretation, and answering the various objections raised against it. It will argue that Bohr's theory is most consistent and coherent when we see that it proceeds from a critical self-examination of the epistemic limitations of our categories of perception and of the concepts of classical physics derived from them. Secondly, it makes the case that Bohr's interpretation of quantum mechanics differs from other scientific theories in that it is, what A.N Whitehead termed a heterogeneous description of nature.

A related matter is that Bohr has been considered a notoriously obscure writer. But there are excellent secondary sources on his interpretation of quantum theory. However, from the point of view of most physicists and philosophers working in quantum foundations, the secondary literature retains aspects of obscurity, vagueness and ambiguity, that seem so prominent in Bohr. The main reason is that Bohr's epistemology is situated in the Continental tradition, and the philosophical distance with the analytical tradition is too great to easily close. In its expository portions, the article will attempt to bridge this gap, by avoiding contentious metaphysical issues that divides the two traditions and building only within mutually shared or connected philosophical terrain. To do this, I employ an unorthodox approach, which among other things, include two thought experiments.

## 2. What is the Copenhagen interpretation?

We know now that there was no single Copenhagen interpretation. (Howard 2004). The main architects of the interpretation; Bohr, Heisenberg, Pauli and Born each had different ideas about what the Copenhagen interpretation meant. But they ironed out their differences among themselves and never expressed it in public in any significant way. But since the prime mover behind the Copenhagen interpretation was Bohr and the others spoke in his name, I will use the term Copenhagen interpretation to refer exclusively to Bohr's views, unless otherwise specified.

Almost parallel with the revival of interest in quantum foundations witnessed recently, but on a far smaller scale, there has been a revival of interest in the thought of Niels Bohr. Since around the mid-nineteen eighties, a number of philosophers and historians of science have been reading Bohr's work very carefully. There is almost universal consensus in the literature on Bohr's thought that has been produced since then that Bohr was not a positivist, instrumentalist or metaphysical anti-realist, at least not in any generic sense of these terms. Influential studies on the life and thought of Bohr by Henry Folse (1985), Jan Faye (1991), Dugald Murdoch (1987) and John Honner (1987) appeared in the 1980s and early 90s. These studies have advanced our understanding of Bohr's thought much further, revealing its complexity, nuance and philosophical depth and the fact that Bohr paid careful attention to the very foundational problems he is accused of ignoring.

While all of these studies do not agree on all the points, there is a certain convergence towards the following themes: 1) Bohr was a metaphysical realist and believed in the existence of objects like electrons etc independent of our knowledge of them. 2) His thinking was clearly influenced directly or indirectly by Kant's transcendental philosophy. 3) Bohr's interpretation does not involve wave collapse and none of his published writings use the term[3]. In the following three decades, there has been an enormous increase in Bohr scholarship, with dozens of scholars following up on the ground-breaking work done in the eighties. While some writers have questioned one or more of the three themes mentioned, others have developed particular themes with ever closer readings of Bohr's texts. For a cross-section about the current debates about how to interpret Bohr, see Beller (1992), Howard (1994), Brock (2003), Perovic (2013) Shomar (2008) Camilleri (2017), Kuark-Leite (2007). My own interpretation of Bohr would develop the Kantian themes in Bohr's work far more explicitly than the commentators who have explored the Kantian strains in Bohr's thought have done. (See Honner (1987), Faye (1991) Kaiser (1992), Chevalley (1994) Brock (2009) Cuffaro (2010))[4]

In trying to give a clear account of Bohr's interpretation, I propose to transpose it from its original epistemological setting of early 20th century philosophy[5], to that of contemporary Analytic philosophy. This is because the majority of philosophers and physicists working in quantum foundations today, work within the broad framework provided by Analytic philosophy. It might be objected that such a transposition would be a mutilation of Bohr's philosophy of physics. While I do not dismiss the substance of such an objection, it would miss the point of the article. I am not attempting here to give an exposition of Bohr's philosophy of quantum theory as it is, but rather to present those aspects of his interpretation that taken together form a coherent and consistent whole that provides answers to current foundational questions in quantum mechanics including the measurement problem. The emphasis of this article is purely philosophical and theoretical and not historical. The central concern that animates this discussion is not to locate the precise views of the historical Bohr, but whether those views can form the basis for an interpretation of quantum mechanics that meets the main desiderata of contemporary attempts to solve foundational questions in quantum mechanics, viz a) Quantum systems exist independently of our knowledge of them b) The theory is internally consistent and empirically valid and c) The theory offers a clear solution to the measurement problem. To put it more simply, we are concerned not with exactly what Bohr said, but what Bohr ought to have said. That said, I am not offering a new interpretation of quantum mechanics using Bohr's Copenhagen interpretation as a point of departure; nor am I putting forward a modified form of the Copenhagen interpretation. The account offered here tallies with Bohr's work, though like any interpretation of a thinker, especially one whose thought is as subtle and difficult as Bohr's, it can be contested by those who understand his work differently. And while I will not offer a historical exegesis of Bohr's views as it lies beyond both the aim and scope of this article, I will try to locate my views with reference to the different interpretations of Bohr's work by philosophers and historians of science.

From a Pragmatised Kantian interpretational perspective, I will try to present the Copenhagen interpretation as realist, and as consistent and coherent in relation to the two main objections often raised against it. Both accrue from Bohr's insistence that the observables of quantum theory like position or momentum can only be properly ascribed to quantum systems in the context of a measurement. And that at least part of the measuring apparatus has to be always described classically This has led to the following objections 1) Bohr does not offer a theory about microscopic systems but only about the outcomes of measurements. Even if Bohr did believe in a world of quantum objects which exist independently of our knowledge of them,

---

[3] Murdoch (1987). Faye (2002).
[4] The work of Bitbol and Osnaghi is an exception.
[5] Bohr's interpretation of quantum mechanics was set in the midst of the discursive cross-currents of neo-Kantianism, Machian phenomenalism and logical positivism

this metaphysical realism is just an empty postulation of entities without physical meaning and Bohr's theory is instrumentalism in disguise. 2) By insisting that the measurement apparatus should be described classically, the Copenhagen interpretation divides the world arbitrarily into two: quantum and macroscopic.

Durr and Lazarovici (2019) give a strong version of the second objection: "Copenhagen interpretation…divides the world into two parts: the macroscopic world is classical, the microcosmos is quantum mechanical. The macroscopic world is understandable; the quantum world is not…. The two worlds are basically separated by a cut, often called the *"Heisenberg cut". There are countless papers on where exactly the cut is to be located, i.e., where exactly the dividing line runs between the microscopic (quantum mechanical) and the macroscopic (classical) world...*

*…We will not say much about the Copenhagen interpretation in this book. Why? Because it just does not make sense. How to see that? The fact that every object in the world consists of atoms (and even smaller units), and atoms are described quantum mechanically. So there can be no such division in principle into classical and quantum mechanical worlds."* The macro-world/micro-world division argument has been made several times, though usually in a less emphatic, though no less categorical manner. (See Rovelli (1996) and Zurek (2003))

## 3. Bohr through Bohm.

The Copenhagen interpretation is not only metaphysically realist, but also theory-realist in the specific sense that physics is considered to be an ontological enquiry into nature, not just a tool for making empirical predictions. (Here, my position is close to that of Folse's views on the knowledge of atomic systems attainable according to Bohr. (Folse 1994.)) Since the aim of the first part of this article is mainly expositional, rather than navigate the reader through the labyrinth of decades of Bohr scholarship, I will take a more unconventional approach for the sake of simplicity. I will start with the views of one of the few contemporaries of Bohr, who achieved a sound understanding of Bohr's interpretation: David Bohm. I will use Bohm's insights into the nature of Bohr's interpretation as a propaedeutic for a more thorough and detailed elucidation of the conceptual structure of the Copenhagen interpretation. Bohm, who developed a rival deterministic hidden variable theory to orthodox quantum mechanics, has a very strong claim to have re-inaugurated the discipline of quantum foundations after debates about foundational issues were considered settled after the EPR debate (1935), in favour of the Copenhagen orthodoxy. It was dissatisfaction with what he understood as Bohr's interpretation that led Bohm to seek an alternative, clearer theory of quantum mechanics in his causal interpretation, which is now known as Bohmian mechanics. What is truly interesting is that Bohm's understanding of Bohr's theory changed over the years; partly as a result of conversations with Bohr, partly due to re-reading Bohr's writings and partly because of insights provided by a postdoctoral researcher working with Bohm. Bohm's mature views on Bohr's Copenhagen interpretation was expressed in a series of interviews he gave to the physicist Maurice Wilkins in 1987.

Bohm realised (probably through his conversations with Bohr) that he had not understood Bohr's views correctly when he published his papers on Bohmian mechanics (1952) and even earlier, when he had written his popular textbook on quantum mechanics (Quantum Theory: 1951) which presented what Bohm then thought were Bohr's views. So, over the years, Bohm re-read Bohr several times, trying to understand exactly what Bohr's views were. Bohm believed he finally understood Bohr's position around 1965, when Donald

Schumacher, a post-doctoral researcher, who was reading Bohr closely, conveyed some crucial insights. Bohm said: *"Schumacher emphasized that Bohr was very careful with his language and was trying to present an unambiguous, a very careful way of using language. He balanced it so much that it's very hard to understand… Bohr was trying to develop a consistent way of putting the facts in quantum mechanics. It's very hard to make it consistent. Somebody like Von Neumann put it by saying, "The wave function represents the reality itself." And then suddenly, when a measurement is made, the wave function has to collapse... So it isn't making sense. When you follow that through, you'll find it's very arbitrary and very unclear what it means. Most physicists use it, but still it's a muddle. Von Neumann's language is a muddle. But most physicists don't realize it because they never analyze it. But you see, Bohr put it consistently."* (Bohm 1987. XI)

It is surely one of the greatest ironies of 20th century physics that the physicist whose theory was most persecuted in the name of Bohr's ideas, was one of the very few contemporary physicists who came closest to understanding them. As Bohm said of the situation prevailing in quantum physical research from the 1930s to the late 1980s: *"[E]verybody plays lip service to Bohr, but nobody knows what he says. People then get brainwashed into saying Bohr is right, but when the time comes to do their physics, they are doing something different. That introduces confusion into physics."* (Ibid) Bohm understood Bohr's philosophical motivations (which had to do with the quantum of action) (Bohr 1963 ATHK) in denying that anything at all can be said about the quantum world outside experimental results. He did not agree with them ultimately, seeing his own causal interpretation as an obvious alternative. However, Bohm saw that Bohr's interpretation was consistent and coherent and avoided the two central problems that confronted all the other versions of the orthodox 'wave collapse interpretations.' One was the arbitrary division of reality into a quantum world and a macroscopic world. The other was the observer dependence of obtaining definite measurement results.

Bohm states explicitly in the interview with Wilkins that though Bohr said the properties of quantum objects cannot be described outside experiments, Bohr believed in the existence of an independent quantum reality. This shows how carefully Bohm read Bohr. Bohr's manner of talking about quantum systems generally and the terminology he used had the unfortunate effect of producing the impression of an ambiguous ontological status for quantum systems. But scattered through his lectures and writing, there are many instances where Bohr categorically states that quantum systems are real physical objects. For instance, Bohr writes:

*"We know now, it is true, that the often expressed scepticism with regard to the reality of atoms was exaggerated; for, indeed, the wonderful development of the art of experimentation has enabled us to study the effects of individual atoms . . .*

*. . . We are aware even of phenomena which with certainty may be assumed to arise from the action of a single atom."* (Bohr. 1963. Page 93 and 103)

Bohm tries to explain how Bohr's interpretation of quantum theory deals with the macro-world/micro-world problem and the measurement problem by using Bohm's own concepts of the world 'manifested' in experience, and the world underlying that experience. The manifested world is the world described by classical physics while the underlying world is the quantum world. *"Bohr emphasizes that the experimental conditions must be described in the classical manifest world... And also, the result is described classically. But the meaning of the result must bring in quantum mechanics. It's only the meaning that makes it interesting.*

*But the question is how are you to do this consistently? Because somehow, starting with a classical structure, you're jumping into a quantum meaning. What he said was that this is a whole. He said that because the quantum is indivisible, you cannot separate the quantum from the classical. The quantum of action is indivisible; therefore you cannot say there is a quantum level and a classical level. That's the sort of thing that Von Neumann said, but it gets into a muddle. So what Bohr says is that the distinction between this quantum and the classical is only a logical one. What that means is that it's not an existential distinction of saying, "This is here, and that's there." ...He says the quantum meaning is only described symbolically by the mathematics, which gives the probability of the result…There's a certain wholeness in this Bohr approach, which is not in the Von Neumann approach where he breaks it up in two where he makes the observer and the observed… (Bohr) says that the experiment as a whole, the observer is out there, but it doesn't matter. He's just looking at the classical manifest level where it doesn't matter what he does, and therefore he could be left out of the picture."* (Ibid)

Before building on Bohm's comments, a brief look at Bohr's work is in order. Part of the difficulty in understanding Bohr's thought is that he based his analysis on language and the conditions of possibility of objective communication of scientific facts. Bohr writes: "*However far the phenomena transcend the scope of classical physical explanation, the account of all evidence must be expressed in classical terms. The argument is simply that by the word 'experiment' we refer to a situation where we can tell others what we have done and what we have learned and that, therefore, the account of the experimental arrangement and of the results of the observations must be expressed in unambiguous language with suitable application of the terminology of classical physics*." (Bohr1958.). Since Planck's quantum of action makes it impossible to distinguish between the part of the measurement apparatus involved in the interaction with the quantum system; and the said quantum system; the quantum system exceeds the possibility of direct description. However, the interaction allows us to describe the experiment results that give us knowledge of the quantum system. But the results have to be described using the concepts of classical physics rather than the more fundamental quantum theory, since the very possibility of communication is grounded in ordinary language from which the concepts of classical physics are extracted and theoretically refined. (See Favhroldt 1994).

The audience Bohr addressed was mainly physicists, so he naturally foregrounded his philosophy of quantum mechanics in those themes that are significant and universally indispensable to scientific practice, such as experiments. This necessarily precluded Bohr from giving systematic exposition to the deeper philosophical implications of his treatment of quantum mechanical description and experiments, and of giving a comprehensive account of the exact nature and form of its epistemological and metaphysical commitments. This is why an interpretative reconstruction of Bohr's thought from a more thoroughly philosophical perspective is required in the first place.

I believe the most clear, consistent, and explanatorily powerful understanding of Bohr's ideas can be achieved through unearthing the Kantian analysis of the conditions of possibility of experience which underlies Bohr's analysis of the conditions of possibility of objective communication through ordinary language. Though there are some significant and crucial differences, my interpretational position comes closest to that of Bitbol and Osnaghi (2013), who also attempt an unqualified Kantian reconstruction of Bohr's interpretation. See also Bitbol (2017). Bohr himself made the connection between the limits of description through language and the limits of experience. Bohr writes: "*However, at the same time as every*

*doubt regarding the reality of atoms has been removed and as we have gained a detailed knowledge of the inner structure of atoms, we have been reminded in an instructive manner of the natural limitations of our forms of perception."* (Bohr 1963 103). The core feature of the Copenhagen interpretation is that properties cannot be attributed to quantum system except in the context of a measurement. From this, the principles of the symbolic and stochastic nature of the wave function; and the indispensability of classical concepts for describing measurement results, follow naturally. Both these can be understood in Kantian terms. *"In his reflective analysis of the structure of our capability to know, however, Bohr did not address, as Kant did, mental faculties such as sensibility and understanding. Instead, he focused on a technological counterpart of sensibility, namely the measuring apparatus, and on an intersubjective counterpart of understanding, that is, language."* (Bitbol and Osnaghi 2013)

**4. How the Copenhagen interpretation is realist and quantum mechanics an ontological enquiry into nature.**

The default metaphysical position of most physicists today is some form of naturalism. Being naturalists, they are also scientific realists[6]. (I would later call this default ontology classical realism to distinguish it from Bohr's pragmatic Kantian realism. But right now, we will stick with the more familiar term realism). In the realist picture, physical objects exist in space and time. And among those physical objects are animate organisms and sentient creatures, which include human beings. All creatures adapt to their environment to survive and are capable of sensing their environment to some extent. Higher order creatures with visual ability are able to form a perceptual picture of the environment and its various features. Human beings are unique among sentient creatures in being able to fully form a cognitive picture of the physical reality they inhabit. Homo Sapiens are also unique in that they seek theoretical knowledge of the physical world they inhabit. Scientific knowledge, including physics, are highly refined knowledge structures formed by Homo Sapiens that captures the physical ontology of the world. In the case of modern physics, it is in terms of fundamental entities and forces and the laws that govern their interaction.

Before going into how Bohr's pragmatic Kantian realism is different from the realist picture presented above, let us take a brief look at the central features of Bohr's Copenhagen interpretation. (I do not include the principle of complementarity, since it is not relevant to the question of realism, and it is shared by all variants of the orthodox interpretation. It is not exclusive to Bohr's, though he was its originator) According to Bohr 1) The wave function is a symbolic description of the quantum system, which gives the probability outcomes for measurement. (Bohr 1963) 2) The experimental context and results can only be described by using classical concepts. 3) The ascription of classical concepts to a quantum system outside an experimental context is ill-defined. 4. The wave function is a complete description of quantum reality in that nothing more can be said about quantum reality than what the wave function describes.

---

[6] Here I use scientific realism to refer to theories of scientific realism which are not ontologically committed to physicalism. Putnam (1982) identifies scientific realism as falling into three classes, of which two a) metaphysical realism and b) convergence realism are instances of the sense in which I am using the term.

Now let us return to our discussion of realism. There is a crucial metaphysical assumption that classical realism makes. It assumes that the concepts through which our scientific theories describe the world, are properties that inhere in the world. Is it possible to question this assumption in the Kantian spirit, but without leaving the philosophical terrain of realism? I believe it is possible, and one way to do it is to look at the issue through the lens of evolution. If one were to transpose the Kantian critique into realist terms, it would be to ask how we know that our anthropological species-specific cognitive concepts apply to the physical universe as such? After all, a bat is in no position to theoretically describe the world. Why should we assume that just because we are capable of such theoretical description, that description should also pertain to the world-in-itself rather than only to the picture of the world that our cognitive facilities allow us to form? Whatever the force of the argument, this should usually remain only a philosophical question.

I argue that the Bohrian perspective which was formed from the 1920s to the 1940s as quantum mechanics took shape, can be understood best as taking the Kantian question seriously, not because of apriori philosophical reasons but because of the scientific facts presented by quantum mechanics. Hence, the use of the term pragmatised Kantian realism. Bohr's theory makes most sense when it is understood as an attempt to provide a coherent and intelligible framework to explain emerging quantum phenomena that made no sense within the existing conceptual framework of our classical notions. The particle/wave duality of photons and matter, discontinuity[7] in quantum phenomena, the implications of the uncertainty principle, the inability to give reality to the wave function etc presented no way of resolution within the conceptual structure of classical physics. (Without resorting to hidden variable interpretations, collapse models etc.). The insight that lies beneath Bohr's analysis of language and objectivity is that such quantum phenomena indicate that the classical concepts through which we understand our experience of the world, are breaking down as we attempt to describe a quantum reality that is beyond our direct experience. Our classical concepts are built to describe our experience of the world. But the physical universe is not bounded by the range or form of our experience of it; it is the cause of that experience. (Assuming realism.) In quantum mechanics, we have reached a point where we find that the categories which we have built to comprehend our experience of the world are not sufficient to describe the constitutive cause (the physical universe in-itself) of that experience itself. (We are able to achieve such a description in the context of a measurement.)[8]

The relation between the concepts of experience and the fundamental concepts of classical physics as Bohr saw it is described by Jan Faye (2002): "*Our pre-scientific practice of understanding our environment is an adaptation to the sense experience of separation, orientation, identification and reidentification over time of physical objects. This pre-scientific experience is grasped in terms of common categories like thing's position and change of position, duration and change of duration, and the relation of cause and effect, terms and principles that are now parts of our common language. These common categories*

---

[7] Writing to Erwin Schrodinger in October 1926, Bohr referred to how the limits of applicability of classical concepts follows from the discontinuity of quantum phenomena. *Bohr wrote "since the definition of every concept or rather every word presupposes the continuity of phenomena and hence becomes ambiguous as soon as this presupposition cannot be upheld."* Works 6.

[8] *"[T]he question was how to use classical concepts in the description of quantum processes, and Bohr's answer…was that…these concepts should no longer be interpreted as referring to 'absolute attributes' or 'intrinsic properties' of quantum objects"*. Chevalley.1994.

*yield the preconditions for objective knowledge, and any description of nature has to use these concepts to be objective. The concepts of classical physics are merely exact specifications of the above categories."*

This picture is not classical realism, but is it realism at all? Yes, clearly so. There are real physical entities like electrons and protons that inhabit the sub-atomic world and whose properties manifest themselves under experimental conditions we create to access them. Just because we are not able to describe quantum systems using classical concepts at all times does not mean we have to assume an indefinite reality. In the Copenhagen interpretation, quantum systems have a definite ontological status[9], an ontic state, at all points in time. From the fact that they are unintelligible and inaccessible to us, it does not mean these states don't exist. As a heuristic device, we can imagine an alien species with completely different sensory and cognitive capacities or an omniscient God who can access and describe these states. Or if we are willing to be Platonists about concepts (Like Karl Popper's world of ideas)[10], we can postulate a set of concepts which can describe the ontic state of quantum systems, but which is completely inaccessible to us.

It might be objected that this is only disguised instrumentalism. That while a world of real quantum objects is postulated, it is empty of content, and the theory is ultimately only about macroscopic measurement results. But such an objection does not hold. Instrumentalism concerns itself only with measurement results. It has nothing whatsoever to say about, and is not concerned with, what these results represent or says about the physical world. For the Copenhagen interpretation on the other hand, experimental results give ontological knowledge of physical reality. Quantum systems may in general be inaccessible to us for description, but an experiment creates the special epistemic conditions where we are able to apply our classical concepts to the quantum system. Unlike instrumentalism, we are not interested in macroscopic experimental results for their own sake. We are interested in them because they tell us something about the underlying quantum reality. (This is what Bohm means when he says of Bohr's interpretation that *"(T)he result is described classically. But the meaning of the result must bring in quantum mechanics. It's only the meaning that makes it interesting."*) An experiment creates the epistemic conditions where we are able to grasp the ontic state of the quantum system in terms of the classical concepts built from our experience of the macro-world. To put it another way, we are able to express the ontic state of the system through our classical concepts in the context of an experimental set up. This is what differentiates the Copenhagen interpretation from instrumentalism and makes it truly realistic. Physics is not about measurement results, but it is an ontological enquiry into physical reality.

The evolution of the state of a quantum system through the Schrodinger wave equation, and the results of a measurement are correlated by Postulate A. Postulate A says that the symbolic

---

[9] It has been a curious fact that even the more insightful commentators on Bohr have shied away from attributing an ontic state to quantum systems, while insisting that Bohr held that quantum systems exist outside experiments. Some have even denied that quantum systems can have an ontic state prior to measurement. This seems to me an unnecessary Kantianisation of Bohr's position, to the extent that quantum systems have a noumenal status. Firstly, Bohr himself talks about atomic objects as they exist outside experimental contexts. That is he makes a distinction between the phenomena manifested in the context of an experimental setup and the quantum systems that are the cause of the phenomena. Secondly, once you grant that a physical object exists, it is incoherent to deny that it has an ontic state.

[10] This would mean Platonising Popper's world of ideas. Though it has an autonomous ontological existence, it consists only of concepts constructed by human minds. *Popper (1972)*

description and representative descriptions of the quantum system are complementary and mutually exclusive. That is, the symbolic description of quantum systems through the wave function, and the measurement results obtained by an experiment (What is called wave collapse in standard quantum mechanics), which represents ontological knowledge of the quantum system, are complementary descriptions. The evolution of a given quantum system through time is ordinarily described by the Schrodinger equation. However, when a measurement is performed, values are obtained for observables; that is a direct description of the quantum system becomes possible in terms of the concepts of classical physics like position, energy and so on. At the moment of measurement, and in the period during which the quantum system retains these values, only the second type of description ie the representational description, is operative. The symbolic description through the wave function remains suspended, as it were.

This would mean that in the Copenhagen interpretation, all talk of the wave function of a measured system, is in the strict sense meaningless. The state of the system is completely given by the values obtained for commuting observables. Once the state of the quantum system no longer corresponds to the observed values (the system time evolves), the Schrodinger equation becomes operative again as the means of description. A symbolic description means it is not to be understood as a literal description that corresponds to the quantum system. In the Copenhagen interpretation, ingredients of the symbolic description like superpositions have no physical meaning. The wave function acts as a heuristic device which helps in the performance of experiments and also exercises a regulative function (Chevalley 1994). It also gives certain kinds of information about the system like range of possible values for observables, expectation values etc.

## 5. Thought Experiment

I will try to illustrate this epistemological scenario with a thought experiment. The thought experiment does not merely have explanatory value: it demonstrates the viability and coherence of the epistemological revolution in quantum mechanics Bohr proposed; specifically, his contention that the properties of classical physics can be applied to quantum systems only in the context of a measurement. While the literature offers plenty of neo-Kantian explanations for Bohr's position at the conceptual level, it has never been unambiguously clear whether these explanations are fully coherent. This explanatory gap arises because of our inability to visualise how a fundamental ontic property of an object can apply only in certain epistemic contexts. At the intuitive level, an object either possesses a particular fundamental property at all times; or not at all. The thought experiment bridges this explanatory gap by providing a visualisation of the epistemological schema developed by Bohr.

Imagine a gigantic tank shaped like a blimp of immense proportions made of some semi-opaque substance. There are several circular windows interspersed at irregular intervals along the tank. Inside the tank, thousands of tiny spherical objects whizz through the space obeying an unknown dynamic. An observer looking from outside will see the objects as only foggy outlines. This is the case whether they look through the surface of the tank or one of the windows. Let us call these tiny spherical objects prions. The unknown dynamic is such that the prion never rotates. Now imagine that for every individual prion, there is a long arrow attached to its surface. The arrow keeps rotating in a vertical plane along the trajectory of the prion. The motion of the arrow circumscribes a circle on this plane if the prion is stationary. However, the rotation is not continuous, but happens through series of discrete rotations of the arrow along the sphere's circumference. The only rule is that a single rotation cannot exceed 90 degrees. So, at any point in time, the arrow will be in one of the four quadrants of the imaginary circle its motion forms around the sphere.

Observers who look through the semi-opaque surface will be able to indistinctly make out some kind of objects moving inside the tank. If they observe carefully, in some circumstances they may be able to obtain fuzzy information about its shape, direction of motion etc. They will not be able to see the arrows or even make out that the prions are spherical. The windows are made of coloured glass which are semi-transparent. The coloured glass makes it impossible to gather more about the prions than is possible by looking through the surface. Each window consists of yellow, green, red and blue quarter circles of the same area, and always in that order. As it moves, every prion keeps emitting an electrical pulse in all directions. On the inside of every window, at its dead center is a small piece of sophisticated electronic equipment designed to detect the pulses from the prions. Whenever a prion passes the vicinity of the window the electronic device detects the pulse. However, an observer who is looking through the window still doesn't see anything through the multi-coloured glass. We will make two assumptions about the hypothetical observer 1) The observer has no knowledge of how the electronic device functions and does not even know what it does or how it got there 2) The observer does not possess the concept of a coordinate system, possibly because she comes from a civilisation that is yet to develop advanced mathematical concepts. That is, she does not possess the mathematical knowledge to construct an abstract coordinate system in 3-dimensional space like a Cartesian coordinate system.

Whenever a prion whose trajectory takes it near the window is in a position where its center aligns in a perpendicular line with the center of the window, the electronic device detects this alignment. It instantaneously lights up the window, making the prion visible to an observer for a brief moment. In that moment the observer will notice whether the arrow appears in the yellow, green, red or blue part of the window. Sometimes the arrow is seen in a blurred way at the bottom or top half of the window. Over time, after several such observations of prions, the observer can conclude that a prion can be in six states. With the arrow seen in these places- yellow pane, green pane, red pane, blue pane, blurred-top, blurred bottom. These descriptions correctly correspond to properties of the prions. Each colour corresponds to different orientations of the arrow along the prion's vertical axis. Yellow corresponds to when the arrow circumscribes an angle of 1 to 90 degrees, green to an angle of 91 to 179 degrees, red to an angle of 181 to 270 degrees and blue to 271 to 360 degrees. The blurred positions occur when the arrow occupies a zero degree (blurred top) and 180 degree (burred bottom) position. In these positions, the arrow falls in the intersection between two different coloured glass panes. The effect of the light from the window is such that when the arrow is seen through the intersection of two colours, it appears blurred. It is important to keep in mind that the observer does not know what properties of the prion the various colours and blurred positions correspond to. This is because she does not possess a coordinate system.

But nonetheless, the categories of description the observer uses captures properties of the system. However, they do not apply to the prions at all when they are not observable through the window with their centers aligning perpendicularly. Except in the condition of being in a specific alignment with one of the windows, yellow, green, red, blue and blurred (top and bottom) has no relationship with the prion.

Two things follow. 1. The categories of description express various states of the object, but are not inherent properties of the object. 2. These categories can only be applied in special epistemic conditions and not universally.

The situation in quantum mechanics is analogous, according to Bohr. The prion represents the quantum system, the window the experimental set-up, and the colour classification, the classical concepts used to describe the results of the experiment. Observing a prion's motion through the semi-opaque surface of the tank is analogous to the description of a quantum system through a wave function which is purely symbolic. These two descriptions are complimentary and mutually exclusive. In this analogy, it is very easy to see why the observer is not an instrumentalist though she cannot describe the properties of the prion at all times.

One could object that in this example the prion actually possesses spatial properties, while in the Copenhagen interpretation, quantum systems do not. However, an analogy can only be constructed within the limitations of what is visualisable. It is impossible to visualise an object without a definite location. However, as mentioned before, we always have the option of being Platonists about concepts. The spatial properties of the prion's arrow in the analogy can be taken to stand in for some property of the quantum system that is describable by a concept that is unintelligible to us.

## 6. The Macro/Quantum division and the solution of the measurement problem

A fundamental objection to the Copenhagen interpretation is that it artificially divides the world into quantum and macro-realms. But Bohr never advocated such an ontological division). *"Bohr considered it a matter of course that from an ontological point of view macroscopic objects are basically quantum mechanical…for Bohr the necessity of using classical concepts has a purely epistemic status: it has to do with our access to the world, by means of macroscopic devices that are described by common language (extended by classical physics)."* (Dieks 2017

Bohr's arguments justifying the macro/quantum division bear directly upon the solution of the measurement problem. Hence, I will deal with both the macro-micro division and the measurement problem together through the application of postulate B. While many authors have shown the nature of the epistemic division between quantum system and macroscopic measurement apparatus to be consistent; it has been done within Bohr's original neo-Kantian epistemological framework. Inconsistencies arise when we try to accommodate it to the framework of scientific realism. Any such accommodation will involve a change to both the framework of scientific realism as well as that of Bohr's epistemology. However, the problem of inconsistency can be taken care of if we generalise Bohr's treatment of the measurement process into an epistemological principle that would apply across all metaphysical frameworks. This is the second postulate I proposed. B) The measurement apparatus is the condition of epistemic possibility of accessing quantum phenomena. This generalisation leaves Bohr's neo-Kantian[11] epistemology intact, while it is also broad enough to be extended into the paradigm of scientific realism. This alters the terms of classical scientific realism in the direction of neo- Kantian realism, while leaving its essential and more general features in place.

---

[11] While I use the term pragmatised Kantianism for what I think to be the most accurate description of Bohr's position, many authors have treated Bohr's relationship with Kant's thought differently. What all these expositions have in common is the idea that while Bohr's thought drew upon the Kantian tradition, it also modified it a great deal. I use the term neo-Kantian to indicate this wider position.

To understand the philosophical justification for Bohr's division of quantum system and a macroscopic apparatus, we have to put aside some of our most intuitive preconceptions regarding realism and view the matter from a different framework as Bohr did. We need not adopt a neo- Kantian epistemology as such. Briefly put, there always exist certain epistemic conditions of possibility for our access to any phenomenon whatsoever. Whether the phenomenon is physical or mental. In the macroscopic world we gain access to physical phenomena through our senses and inner cognitive apparatus. But in quantum mechanics, some of these epistemic conditions of possibility happen to be external to our senses and cognitive apparatus. That is, the measuring apparatus is part of the epistemic conditions of possibility of our access to microscopic phenomena. It is thus also the condition of possibility of describing these phenomena. When the status of the measurement apparatus is understood in such a perspective, the division between quantum and classically described worlds is revealed as a purely epistemic distinction.

Let us explain the macro-micro world division through an analogy. Imagine a system which consists of a floor with n number of tiles of equal dimensions but varying colours. The tiles are fitted with a light source and connected to a power source in such a way that each tile can change colour. Each tile has been assigned a number from 1 to n by which one can immediately know the position of the tile within the system. At a fixed interval of a few seconds, all the tiles change their colour simultaneously. There is an observer in the room, whose job it is to record the evolution of the system. The observer possesses a photographic memory and can track the evolution of the system in time by just looking at the tiles. She does not need to make a physical record.

Now we posit three conditions as binding on the system and observer. The first is that the observer always stands feet together on a single tile, covering the complete surface area of that particular tile. The second condition is that the time interval within which the state of the system changes (all the tiles change colours), coincides exactly with the time required for the observer to survey all the tiles from this bodily orientation without moving her position. In other words, when the tiles in the room changes from one set of colours to the next, the observer has exactly enough time to commit the number and corresponding colour of each tile to memory before the colours of the tiles change again. The consequence of this is that the observer doesn't get the time to shift her position and observe the colour of the tile she is standing on. The third condition is that the observer can choose any tile to stand on to observe the room.

The observer can specify the state of the system using just two parameters. The number of the tile and its colour. Let the set of all colours possible for the tiles be C and all members of the set be represented by c. Let the tile the observer is standing on be TS. At any particular point in time the state of the system is completely given by the Set S ( $T_1c, T_2c, T_3c\ldots T_nc$). But the colour of TS is unknown. In effect, this means that we can never obtain a complete description of the state of the system as long as the observer is within the system. Only a view from outside can give a complete description of the state of the room. As long as the observer is within the room, the description of the room will be incomplete. The description by the observer is given by the Set S\TS where $T_1 \leq TS \leq T_n$. In other words, the method of description can be applied to any tile in the room. But it cannot be applied to all tiles at the same time. The colour and number of the tile that the observer stands on, TSc, can be described through the same means, provided the observer is standing on a different tile. But in that case, the second tile will have to be left out of the description of the total system. The method of description in this case is a primitive one, but the same principle can apply to any kind of description, whether they involve more complex methods or use theoretical models. The cardinal feature of a scenario like this is that while a uniform method of description, or a single unified theory can describe the entirety of the system in principle, a sub-system which is the vantage point from which the description is conducted, has to be excluded from that description in practice.

Consider the analogy between quantum mechanics and the above thought experiment. The floor represents the universe and the tiles various systems within the universe. As our fundamental theory is quantum mechanics (analogous to 'the see and memorise' primitive description in the experiment), everything in the universe including macroscopic systems can be described by it. But the process of measurement is the very condition of epistemic possibility for our conscious minds to access the microscopic world. (We are talking of the mind/external world distinction here in a purely epistemological sense. The mind can be the product of purely physical processes and considered to supervene on the physical world, be an epiphenomenon or an emergent phenomenon.) As our preferred epistemic vantage point to describe the microscopic world, we are unable to include the measurement device within the quantum description, just as in the case of the woman standing on the tile in our thought experiment. To sum up the Copenhagen view, all systems can be described quantum mechanically, but we cannot describe all systems quantum mechanically at the same time. There will always be a privileged epistemic vantage point, which excludes itself from the description of the total system, in the very process of making that description. This is what Bohm meant when he said: "*So what Bohr says is that the distinction between this quantum and the classical is only a logical one. What that means is that it's not an existential distinction of saying, "This is here, and that's there*". In other words, Bohm correctly points out that the distinction between the classical system and the quantum system is only epistemological, not ontological. If it were ontological, it would make no sense, as Durr and Lazorovici states.

Let us examine Postulate B more closely. This is required to deal with the measurement problem and will also provide further philosophical support for the explanation offered here for the macro-micro world division. What exactly is meant by an 'epistemic condition of possibility' in physical science? In biology we are able to see a cell or a microbe only through a microscope. In astronomy, we are able to view distant astrophysical objects only through instruments like the telescope. These technical instruments are a necessary condition for our access to microbiological or astrophysical objects. What is unique about the situation in quantum mechanics where quantum phenomena are accessible only through an experimental set up?

The relevant distinction here is that there is a difference between saying x is our means of access to a phenomenon, and saying that x is the epistemological condition of possibility of our access to a phenomenon. The difference is that in the former case, say of a microscope used to magnify the cell's image; we are able to describe the process by which the microscope creates the magnification effect. That is, the process of how an instrument gives us access to a phenomenon can be described. But we cannot describe why a macroscopic measurement apparatus gives us the kind of access to a quantum phenomenon that enables us to ascribe classical properties to it. It is a brute fact and cannot be analysed further. This is why it is termed an epistemic condition of possibility.

This explanation might still sound obscure, or as an adhoc philosophical addition meant to paper over the consequences of the classical-quantum divide. But the reason it may appear this way is because we have taken the realist picture of the universe for granted to such an extent that we have forgotten the epistemological route by which we got there. This is not to suggest that the realist picture of the universe is wrong. The realism anti-realism debate need not concern us. What concerns us is that there has been a philosophical forgetting among physicists and philosophers of physics. Even if we subscribe to a physicalist account of the universe, the fact remains that what is available to us directly is only our subjective conscious experience. That experience may at the ontological level be the result of the neurochemical activity in the brain as the physicalist says, but that does not change the fact that epistemologically our enquiry into the nature of the world can start only with what we have: the world as given to our conscious experience. We have to provide a rational account of how we theoretically travel from the object given in our conscious experience to the physical object that causes that experience.

Scientific investigation starts with descriptions of everyday objects given to us in conscious experience. These objects are immanent in our consciousness. To arrive at scientific realism, we postulate a universe of physical objects external to our consciousness, which acts on our senses to produce the subjective experience of the objects immanent in our consciousness. The table that I perceive in my consciousness is the result of the interaction of my senses with a physical table that exists in the physical universe. My description of the table (whether it is through classical physics, quantum mechanics, or some other fundamental theory to be discovered in the future), is centered around the table that is within my consciousness. But I believe that the object in my consciousness represents information about its counterpart in the physical universe outside my consciousness. That is, through my description of the object in my consciousness, I indirectly describe the physical object; so that for all practical purposes I can ignore the mediating entity. That is, ignore the representational object given in my consciousness that provides access to the physical entity. I can proceed as if I am directly describing the physical system situated in the physical universe. Thus, experience is abstracted out of the theoretical description of nature. This is what A.N Whitehead called thinking homogeneously about nature. (Whitehead 1964) That is, thinking about nature without reference to the fact that we are thinking about nature.

Part of why Bohr's explanation of the role of the measuring apparatus is so widely misunderstood is that he is referring back to what has been forgotten in a kind of philosophical amnesia. Namely, that we can treat the physical universe as a self-sufficient closed domain only after we have built an epistemological bridge from our pre-theoretical subjective conscious experience of the world to the ontologically objective physical world that lies behind that experience. It is only by standing on this bridge that we can rationally account for our access to the physical world. If we try to do without it, science and reason must part ways.  The reason for the philosophical forgetting of the pre-theoretical level is due to a certain structural isomorphism between our common sense experience of the world and the descriptions of classical physics. *"In so far as one considers classical physics, the indispensability of the pre- theoretical level can go unnoticed, since the pretheoretical treatment of the measurement process can be made isomorphic to its theoretical description. So much so that isomorphism could in this case be conflated with identity."* Bitbol and Osnaghi *(2013)*

When Bohr talks about experience, about our collective agreement about measurement results as the only thing we can say about quantum systems, he is bringing back into the framework of the theoretical description, the world of experience we had abstracted out of it in classical physics. He does this because he thinks that unlike in the case of classical mechanics where we can treat the physical universe, to use Whitehead's terms 'homogeneously', we can account for the results of quantum mechanics only by a 'heterogeneous' description. That is we think about the physical universe with reference to the fact that we are thinking about it. Heisenberg, agreed with him. *"'In classical physics science started from the belief — or should one say from the illusion? —that we could describe the world or at least part of the world without any reference to ourselves. This is actually possible to a large extent…It may be said that classical physics is just that idealization in which we can speak about parts of the world without any reference to ourselves…Does the Copenhagen interpretation of quantum theory still comply with this ideal? One may perhaps say that quantum theory corresponds to this ideal as far as possible. Certainly quantum theory does not contain genuine subjective features, it does not introduce the mind of the physicist as a part of the atomic event. But it starts from the division of the world into the "object" and the rest of the world, and from the fact that at least for the rest of the world we use the classical concepts in our description…the use of the classical concepts is finally a consequence of the general human way of thinking. But this is already a reference to ourselves."* (Heisenberg 1958. See also Hooker 1994). Once we have understood this, it is easier to appreciate Bohr's position by placing it within its proper setting; that of a heterogeneous theoretical description of physical reality.

Take the scenario where we perceive any part of our own body by looking at it. The picture scientific realism presents is that there is a physical system BD that constitutes a part of the human body. Light rays reflecting off BD impacts another physical system E which constitutes the eye of the body. Other physical systems including nerves, carry the stimulus to the physical system BR, the Brain, which produces consciousness and the experience of the body. Now the object experienced as the body by my ego is not BD, but an object given in the field of my consciousness. This object has been created by the brain according to the information it extracted and processed from the external world. Everything that we experience is a creation of the brain. (I am using a physicalist framework here because what applies to physicalism in this case, applies a fortiori to any kind of dualism of matter and mind.) This is why the neurobiologist Anil Seth used the by now popular phrase 'The brain hallucinates your conscious reality'. Seth's formulation is useful in so far as it emphasises that everything we experience, including our experience of the external world is a recreation of consciousness, analogous to how a computer recreates a movie from the information encoded electronically. (In the case of the brain the information is encoded neuro-chemically.) But the phrase is also misleading. A hallucination bears no relationship to reality. But as far as scientific realism is concerned, the body I see in my experience represents information about my physical body BD, which exists in the physical world.

So whether we take the route of traditional epistemology (In the tradition that runs from Descartes to Husserl), or that of scientific realism, we are confronted with the same scenario. Whatever our scientific theory may be, the direct referent of our description, is an object perceived within consciousness. Now how do we, within the field of our consciousness created by the brain, perceive the objects that appear exterior to my selfhood, that is, to the interiority of my mental life? The simple answer is that we do it through our senses. When I experience opening my eyes or orienting my line of vision with an object in terms of angle and distance, I perceive an object. Now this entire process, as experience, takes place in the theatre of consciousness.[12] It is from this point that we have to build an epistemological

---

[12] We believe the objects given to us in consciousness as objects-in-the-world exist in a permanent way outside

account that takes us to the object's physical counterpart in the world which physics seeks to study.

Now we come to the measurement problem. To repeat, in the case of any phenomenon, we have to assume certain epistemic conditions of possibility. For the theoretical descriptions of classical physics, our senses and cognitive apparatus constitute those primary conditions. But in the case of quantum mechanics, some of the epistemic conditions of possibility of accessing microscopic phenomena are external to our cognitive apparatus and senses. That is, it now includes the measuring apparatus. The easiest way to think about this, metaphorically, is to think of the measuring apparatus as an extension of our senses. From this it naturally flows that the measuring apparatus itself can never be a part of the description of any quantum system. The epistemic condition of possibility of a phenomenon and the phenomenon itself can never be put in the same descriptive frame. In the simplest of examples, we see things through our eyes, but we can never see our eyes themselves directly. So, at the ontological level, the measurement apparatus which is a macro system, is constituted by sub-atomic entities and is in principle describable by the Schrodinger wave function. But at the epistemological level, being among the conditions of possibility of describing the microscopic world, the measuring apparatus has to be left out of the description. "What preconditions the possibility of a quantum description cannot be described quantum-mechanically in the very process of describing." (Bitbol and Osnaghi 2013). It can be described quantum mechanically (at least in principle), but only through another measurement. And that measurement device will then have to be excluded from the description.

The Kantian framework I have adopted to interpret Bohr, is similar to that presented by Bitbol and Osnaghi (2013). But there are several differences. The most significant are a) They present the elements of Bohr's thinking, from which Postulates A and B can emerge, but they do not attempt such an epistemological generalisation. b) They don't read Bohr as committed to realism as unambiguously as I do here; nor in as strong a form. C) Though they suggest strongly that Bohr overcomes the measurement problem, they take a more conservative stance and leave the issue relatively open ended. Zinckernagel (2016) argues that Bohr 'dissolves' the measurement problem by insisting on a classical description of the apparatus; thus precluding the possibility of the pointer being in a superposition ab initio. However, he holds that Bohr was not a quantum fundamentalist. Quantum fundamentalism is the belief that "Everything in the universe (if not the universe as a whole) is fundamentally of a quantum nature." Zinckernagel (2015). Apart from that I think this is a misreading of Bohr, this opens up the problem of the macro-micro world division. Zinckernagel argues that it can be overcome by understanding Bohr as holding to "ontological contextualism", where an object is quantum or classical according to the context. Within the domain of physics, which deals with fundamental ontology, the idea of an ontology that depends on a context is simply not coherent.

A clear solution to the measurement problem emerges naturally from the main discussions in the article. The physical reality of quantum systems, in conjunction with postulates A and B, dissolves the measurement problem. The measurement problem is presented in several forms. The two most significant are the problem of collapse, and the problem of definite outcomes. If quantum objects are always described through a linear wave equation, they are in a

---

the interiority of my mental life because a) The nature of these objects cannot be altered by thought and b) Other people perceive the same object in similar cognitive conditions.

superposition of eigenstates of a relevant observable operator. But when a quantum system interacts with a measurement apparatus, the pointer is not found in a superposition of different outcomes, but has a definite position. This is the problem of definite outcomes. The problem of definite outcomes arises from the mistake of giving literal meaning to a symbolic description. Superpositions are abstract entities in a mathematical formalism, with no empirical meaning. They can be interpreted empirically only in terms of probabilities for possible measurement outcomes. Bohr insisted that the results of a measurement, including pointer position, can only be described classically. Since the form of our perception allows objects to be represented only with definite spatial locations, there is no mystery to the pointer being in a definite position. As Folse (1994) writes, *"The cat in Schrodinger's famous experiment is quite dead or quite alive long before any observer opens the box and looks... What would be ill-formed, of course, would be a statement attempting to predicate a state superposing both being alive and being dead to the cat."*

The problem of collapse is the inconsistency between the unitary evolution of the state of a quantum system and a random discontinuous 'collapse' of the wave function into definite values for observables on measurement. *"Textbook quantum theory provides two rules for the evolution of the wave function…A deterministic dynamics given by Schrödinger's equation when the system is not being "measured" or observed, and a random collapse of the wave function to an eigenstate of the "measured observable" when it is. However…quantum theory does not explain how to reconcile these two apparently incompatible rules."* (Goldstein 2021).

For Bohr, there is no conflict between a deterministic description and a rule of random collapse. This is because they are not the same type of description. The evolution of the quantum system through the Schrodinger equation is a purely symbolic description that does not describe the properties of the system. A measurement however, is a representational description which captures the ontic state of the quantum system in terms of the values of observables. They cannot be applied at the same time because one cannot have an incomplete description and a complete ontic description at the same time. They necessarily exclude each other and can work only as complementary descriptions. The state of the system at the time of measurement is not given by the collapsed wave function, but the set of values of the measured observable and its commuting observables. Dieks (2017 ) rightly notes that Bohr probably *"did not think of the collapse of the wave function as a physical process, as a rival in the formalism of unitary Schrodinger evolution. From a modern perspective, this places Bohr in the camp of non-collapse interpretations of Quantum mechanics."*

However, it has to be noted that it is the traditional formulation of the measurement problem that the Copenhagen interpretation succeeds in solving. It would be premature to be confident that we have understood all the dimensions of the measurement problem. Further philosophical analysis, or theoretical developments in physics can reveal aspects of the measurement problem that are as yet unknown. In such a case, one would have to revisit whether Bohr's epistemological fortifications hold out against fresh challenges.

**References**

Barret, Jefferey, A. 2019. Conceptual Foundations of Quantum Mechanics. Oxford University Press. Oxford.

Bacciagaluppi, Guido. 2013. Measurement and Classical Regime. The Oxford Handbook of Philosophy of Physics. Chapter 12. Oxford.

Beller, Mara.1992. The Birth of Bohr's Complementarity: The Context and the Dialogues. Studies in History and Philosophy of Science, 23: 147–180.

Bitbol Michel, Osnaghi Stefano. 2013. Bohr's complementarity and Kant's epistemology. Séminaire Poincaré 17: 145–166.

Bitbol. Michel. 2017. "On Bohr's Transcendental Research Program", in Faye and Folse (eds.), Niels Bohr and the Philosophy of Physics: Twenty First century perspectives. London, Bloomsbury Academic. 47–66.

Bohm, David. 1951. Quantum Theory. Prentice-Hall. New York.

Bohm, David. 1952. A Suggested Interpretation of the Quantum Theory in Terms of "Hidden" Variables. I and II, Physical Review. 85.2 85-156 and 85-180.

Bohm, David. (1987). Session XI. https://www.aip.org/history-programs/niels-bohr- library/oral-histories/32977-11

Bohr, Niels. 1963. Atomic Theory and the Description of Nature. Cambridge University Press. London.

Bohr, Niels.1958. Atomic Physics and Human Knowledge. 39. John Wiley. New York.

Bohr, Niels. 1963. Essays 1958-1962 on Atomic Theory and Human Knowledge.5. 93.103.Interscience Publishers. New York.

Bohr. Niels. 1985.Collected Works, Vol. 6: Foundations of Quantum Mechanics I (1926– 1932), J. Kalckar ed. North- Holland.462. Amsterdam.

Brock, Steen. 2003. Niels Bohr's Philosophy of Quantum Physics in the Light of the Helmholtzian Tradition of Theoretical Physics, Berlin: Logos Verlag.

Brock, Steen. 2009. Old Wine Enriched in New Bottles: Flavours of Kantianism in Bohr's Complimentarity. Constituting Objectivity: Transcendental Perspectives on Modern Physics. Bitbol, Michel, Kerzberg Pierre and Petiot, Jean (eds). 301-316.

Camilleri, Kristian. 2017. Why Do We Find Bohr Obscure? Reading Bohr as a Philosopher of Experiment. J. Faye and H. Folse (eds.) Niels Bohr and the Philosophy of Physics. Twenty First Century Perspectives. London. Bloomsbury Academic. 19-46.

Cuffaro, Michael E. 2010. The Kantian Framework of Complementarity. Studies in History and Philosophy of Modern Physics, 41 (2010), 309-317

Chevalley, Catherine. 1994. Bohr's words and the Atlantis of Kantianism. Niels Bohr and Contemporary Philosophy. J. Faye and H. Folse (eds.). 33-56. Springer Science+Business Media Dordrecht.

Dieks, Dennis. 2017. Niels Bohr and the Mathematical Formalism. J. Faye and H. Folse (eds) Niels Bohr and the Philosophy of Physics. Twenty First Century Perspectives. London. Bloomsbury Academic . 303.

Durr and Lazarovici. 2019. Understanding Quantum Mecahnics: The World According to Quantum Foundations. vi to vii Springer

Favrholdt, David. 1994. Niels Bohr and Realism. Niels Bohr and Contemporary Philosophy. J. Faye and H. Folse (eds.) 77 to 96. Springer Science+Business Media Dordrecht.

Faye, Jan 1991. Niels Bohr: His Heritage and Legacy. An Anti-realist View of Quantum Mechanics. Springer Science + Business Media, BV. Dordechtd.

Faye, Jan, 2002. Copenhagen Interpretation of Quantum Mechanics, The Stanford Encyclopedia of Philosophy (Winter 2019 Edition), Edward N. Zalta (ed.), URL = <https://plato.stanford.edu/archives/win2019/entries/qm-copenhagen/>.

Fosle, Henry, 1985. The Philosophy of Niels Bohr: The Framework of Complementarity, Amsterdam: North Holland.

Folse, Henry 1994. Bohr's Framework of Complementarity and the Realism Debate. Niels Bohr and Contemporary Philosophy. J. Faye and H. Folse (eds). 119 to 140. Springer Science+Business Media Dordrecht.

Goldstein, Sheldon 2021. Bohmian Mechanics', The Stanford Encyclopedia of Philosophy (Fall 2021 Edition), Edward N. Zalta (ed.), URL=<https://plato.stanford.edu/archives/fall2021/entries/qm-bohm/>.

Harrigan N and Spekkens R. Einstein, incompleteness, and the epistemic view of quantum states. Foundations of Physics. 40 (2): 125.

Heisenberg, Werner, 1958. Physics and Philosophy: The Revolution in Modern Science. New York. Harper and Brother Publishers. 55-56

Hooker, Clifford.1994. Bohr and the Crisis of Empirical Intelligibility: An Essay on the Depth of Bohr's Thought and Our Philosophical Ignorance. Niels Bohr and Contemporary Philosophy. J.Faye and H. Folse (eds). 155-199. Springer Science+Business Media Dordrecht.

Howard, Don. 1994. What makes a concept classical? Towards a reconstruction of Niels Bohr's philosophy of physics. Niels Bohr and Contemporary Philosophy. J. Faye and H. Folse (eds.) 201-230. Springer Science+Business Media. Dordrecht.

Howard, Don 2004. 'Who invented the 'Copenhagen interpretation'? A study in mythology. Philosophy of Science 71: 669– 82.

Honner, John 1987. The Description of Nature: Niels Bohr and The Philosophy of Quantum Physics. Oxford: Clarendon Press.

Kaiser, David. 1992. More Roots of Complementarity: Kantian Aspects and Influences, Studies in History and Philosophy of Science. 23: 213–239.

Kauark-Leite, Patricia. 2017. Transcendental versus Quantitative Meaning of Bohr's Complementarity Principle. J. Faye and H. Folse (eds.), Niels Bohr and the Philosophy of Physics,

Twenty First Century Perspectives. London: Bloomsbury Academic. 67-90.

Murdoch, Dugald. 1987. Niel's Bohr's Philosophy of Physics. Cambridge University Press. London.

Oldofredi, Andrea and Esfield, Michel. 2019. Observability, Unobservabilty and the Copenhagen interpretation in Dirac's methodology of physics. http://philsci-archive.pitt.edu/16618/1/Unobservability_Dirac_Quanta.pdf.

Putnam, Hilary. 1982. Three Kinds of Scientific Realism. The Philosophical Quarterly, 32, 128. 195-200.

Perovic, Slobodan. 2013. Emergence of Complementarity and the Baconian roots of Niels Bohr's Method. Studies in History and Philosophy of Science Part B: Studies in History and Philosophy of Modern Physics 44(3): 162–173.

Popper, Karl 1972. Objective Knowledge: An Evolutionary Approach. Oxford University Press. Oxford.

Rovelli, Carlo. Relational quantum mechanics. International Journal of Theoretical Physics 35, 1637–1678 (1996).

Shomar Towfic, 2008. Bohr as a Phenomenological Realist. Journal for General Philosophy of Science, 39: 321– 349.

Whitehead, Afred Northop. 1964. The Concept of Nature. Cambridge University Press. Cambridge.

Zinckernagel, Henrick . 2015. Are we living in a quantum world? Bohr and quantum fundamentalism.

F. Aaserud and H. Kragh (eds.). One hundred years of the Bohr atom: Proceedings from a conference (Scientia Danica. Series M: Mathematica et physica, Volume 1), Copenhagen: Royal Danish Academy of Sciences and Letters, 419–434.

Zinckernagel Henrick. 2016. Niels Bohr on the wave function and the classical/quantum divide. Studies in History and Philosophy of Modern Physics, 53: 9–19.

Zurek, H. Wojciech. 2003. Decoherence, einselection, and the quantum origins of the classical. Reviews of Modern Physics. 75, 715–775.